\documentclass[a4paper,11pt]{article}

\usepackage{jheppub_mod} 
\usepackage[T1]{fontenc} 

\usepackage{hyperref}
\usepackage{graphicx}
\usepackage{amsmath,amssymb,slashed}
\usepackage{booktabs,tabulary}
\usepackage{dsfont}
\usepackage{color}
\usepackage{multirow}
\usepackage{subcaption}

\newcommand{\mpi}{M_{\pi}}
\newcommand{\mpii}{M_{\pi^0}}
\newcommand{\Order}{\mathcal{O}}
\newcommand{\keV}{\,\text{keV}}
\newcommand{\MeV}{\,\text{MeV}}
\newcommand{\GeV}{\,\text{GeV}}

\newcommand{\beq}{\begin{equation}}
\newcommand{\eeq}{\end{equation}}
\newcommand{\diff}{\text{d}}

\newcommand{\sthr}{s_\text{thr}}

\newcommand{\PiO}{\pi^0}
\newcommand{\F}{\mathcal{F}}
\newcommand{\A}{\mathcal{A}}
\newcommand{\G}{\mathcal{G}}
\newcommand{\disc}{\text{disc}\,}

\renewcommand{\Im}{\text{Im}\,}

\newcommand{\Mw}{M_\omega}
\newcommand{\Mphi}{M_\phi}
\newcommand{\gwg}{g_{\omega\gamma}}
\newcommand{\Gw}{\Gamma_\omega}
\newcommand{\Gphi}{\Gamma_\phi}
\newcommand{\Mwb}{\bar M_\omega}
\newcommand{\Gwb}{\bar\Gamma_\omega}

\newcommand{\Gr}{\Gamma_\rho}
\newcommand{\grg}{g_{\rho\gamma}}

\newcommand{\eps}{\epsilon}

\allowdisplaybreaks[1]

\preprint{INT-PUB-19-030}
\title{Three-pion contribution to hadronic vacuum polarization}

\author[a]{Martin Hoferichter,}
\author[b]{Bai-Long Hoid,}
\author[b]{and Bastian Kubis}

\affiliation[a]{
Institute for Nuclear Theory, University of Washington, Seattle, WA 98195-1550, USA}

\affiliation[b]{
Helmholtz-Institut f\"ur Strahlen- und Kernphysik (Theorie) and \\
Bethe Center for Theoretical Physics, Universit\"at Bonn, 53115 Bonn, Germany}

\emailAdd{mhofer@uw.edu}
\emailAdd{longbai@hiskp.uni-bonn.de}
\emailAdd{kubis@hiskp.uni-bonn.de}

\abstract{We address the contribution of the $3\pi$ channel to hadronic vacuum polarization (HVP) using 
a dispersive representation of the $e^+e^-\to 3\pi$ amplitude. This channel gives the second-largest
individual contribution to the total HVP integral in the anomalous magnetic moment of the muon $(g-2)_\mu$, 
both to its absolute value and uncertainty. It is largely dominated by the narrow resonances $\omega$ and $\phi$,
but not to the extent that the off-peak regions were negligible, so that 
at the level of accuracy relevant for $(g-2)_\mu$ an analysis of the available data as model independent as possible 
becomes critical. Here, we provide such an analysis
based on a global fit function using analyticity and unitarity 
of the underlying $\gamma^*\to3\pi$ amplitude and its normalization from a chiral low-energy theorem, 
which, in particular,  allows us to check the internal consistency 
of the various $e^+e^-\to 3\pi$ data sets. 
Overall, we obtain
$a_\mu^{3\pi}|_{\leq 1.8\GeV}=46.2(6)(6)\times 10^{-10}$
as our best estimate for the total $3\pi$ contribution consistent with all (low-energy) constraints from QCD. 
In combination with a recent dispersive analysis imposing the same constraints on the $2\pi$ channel below $1\GeV$, this covers nearly $80\%$ of the 
total HVP contribution, leading to
$a_\mu^\text{HVP}=692.3(3.3)\times 10^{-10}$
when the remainder is taken from the literature,
and thus reaffirming the $(g-2)_\mu$ anomaly at the level of at least $3.4\sigma$.
As side products, we find for the vacuum-polarization-subtracted masses $\Mw=782.63(3)(1)\MeV$ and $\Mphi=1019.20(2)(1)\MeV$, 
confirming the tension to the $\omega$ mass as extracted from the $2\pi$ channel. 
}

\begin{document} 
\maketitle

\section{Introduction}
\label{sec:intro}

Three-particle decays subject to strong final-state interactions are notoriously difficult to
describe in a fully model-independent way, i.e., without assumptions on intermediate states of the decay 
or other approximations of the hadron dynamics. 
One of the simplest examples is the three-pion decay of vector mesons, $V=\omega,\phi$, which
phenomenologically is dominated by the $\rho(770)$ resonance formed
in the final-state rescattering of the pions. 
However, a description beyond a simple isobar model is challenging, especially given that the decay is out of reach for
low-energy effective field theories.  
A strategy to control the pion final-state interactions based on analyticity and unitarity 
was first developed in the context of $K\to 3\pi$~\cite{Khuri:1960zz} and applied to
$\omega\to3\pi$ as early as~\cite{Aitchison:1977ej}. These Khuri--Treiman (KT) equations 
have since become a standard tool in three-particle decays, 
with recent applications specifically to $\omega,\phi\to3\pi$ decays in~\cite{Niecknig:2012sj,Danilkin:2014cra,Dax:2018rvs},
in part triggered by significant progress in the determination of the $\pi\pi$ phase shifts that are required as 
crucial input in the solution~\cite{GarciaMartin:2011cn,Caprini:2011ky}. 

A detailed, model-independent understanding of hadronic amplitudes can have significant impact beyond low-energy QCD itself, most notably in low-energy searches
for physics beyond the Standard Model (SM) such as the anomalous magnetic moment of the muon $a_\mu=(g-2)_\mu/2$, whose SM prediction currently disagrees with experiment~\cite{Bennett:2006fi}
\beq
a_\mu^\text{exp}=116\ 592\ 089(63)\times 10^{-11}
\eeq
at the level of $3$--$4\sigma$. To confront the SM with upcoming experiments at Fermilab~\cite{Grange:2015fou} and J-PARC~\cite{Abe:2019thb}, one needs to be able to control 
the theoretical uncertainties at a commensurate level. In this context, the issue of hadronic modeling is most severe in the hadronic light-by-light (HLbL) contribution, 
for which a data-driven dispersive approach has only been recently 
developed~\cite{Hoferichter:2013ama,Colangelo:2014dfa,Colangelo:2014pva,Colangelo:2015ama,Colangelo:2017qdm,Colangelo:2017fiz,Hoferichter:2018dmo,Hoferichter:2018kwz}.\footnote{See~\cite{Blum:2016lnc,Blum:2017cer,Gerardin:2019vio} for recent progress in lattice-QCD calculations of HLbL scattering.}
In contrast, the leading hadronic contribution, hadronic vacuum polarization (HVP), is, in principle, fully determined by 
the cross section for $e^+e^-\to\text{hadrons}$~\cite{Bouchiat:1961,Brodsky:1967sr}, 
and indeed a combination of the analysis of exclusive channels, inclusive data, and perturbative-QCD constraints
are used for current estimates of the HVP contribution~\cite{Davier:2017zfy,Keshavarzi:2018mgv,Jegerlehner:2018gjd,Benayoun:2019zwh}.\footnote{This does not
apply to space-like approaches, as in lattice QCD~\cite{Borsanyi:2017zdw,Blum:2018mom,Giusti:2018mdh,Shintani:2019wai,Davies:2019efs,Gerardin:2019rua,Aubin:2019usy} or the MUonE proposal~\cite{Abbiendi:2016xup}, which are complementary
to but not yet competitive with the time-like approach.} 
However, only the compilations from~\cite{Davier:2017zfy,Keshavarzi:2018mgv} are exclusively based on the direct integration of the data, while~\cite{Jegerlehner:2018gjd,Benayoun:2019zwh} do involve some
model assumptions, in particular for the $\omega$ and $\phi$ contributions. In general, tensions among data sets are typically taken into account by a local error inflation.  

For the lowest-multiplicity channels that dominate HVP at low energies the available constraints from analyticity and unitarity (as well as low-energy theorems) are powerful enough
to define a global fit function that the data need to follow if consistent with all QCD constraints. Such an approach to the $2\pi$ channel below $1\GeV$ has recently been completed
in~\cite{Colangelo:2018mtw}, relying on a close relation between $\pi\pi$ scattering, the pion vector form factor, and the HVP integral~\cite{DeTroconiz:2001rip,Leutwyler:2002hm,Colangelo:2003yw,deTroconiz:2004yzs,Ananthanarayan:2018nyx}. As a result, it was found 
that despite the known tension between BaBar~\cite{Aubert:2009ad,Lees:2012cj} and KLOE~\cite{Ambrosino:2008aa,Ambrosino:2010bv,Babusci:2012rp,Anastasi:2017eio}, 
each data set by itself is consistent with QCD constraints, and a global fit then defines an average that only uses as additional input information on the covariance matrices as provided by experiment. 

Here, we extend this strategy to the $3\pi$ channel, which produces both the second-largest contribution to the total HVP value and its uncertainty.
Instead of the pion vector form factor, 
the underlying hadronic amplitude becomes $\gamma^*\to3\pi$, which was studied in detail in the context of the pion-pole
contribution to HLbL scattering~\cite{Schneider:2012ez,Hoferichter:2012pm,Hoferichter:2014vra,Hoferichter:2017ftn,Hoferichter:2018dmo,Hoferichter:2018kwz}
and further emerges in the two-pion contributions via the left-hand cut in $\gamma^*\gamma^*\to\pi\pi$~\cite{GarciaMartin:2010cw,Hoferichter:2011wk,Moussallam:2013una,Danilkin:2018qfn,Hoferichter:2019nlq}.
For an isoscalar-photon virtuality $q^2=M_\omega^2,M_\phi^2$, this amplitude is directly related to the three-particle decays of $\omega$ and $\phi$, and indeed
the KT approach can be generalized to obtain a dispersive representation of the $e^+e^-\to3\pi$ cross section, see Sect.~\ref{sec:g3pi} for a short review. 
The $\omega$ and $\phi$ resonance peaks thus constitute the most conspicuous features of the cross section, but for the HVP integral also the off-peak regions 
need to be controlled, with QCD determining, via 
the Wess--Zumino--Witten anomaly~\cite{Wess:1971yu,Witten:1983tw}, the normalization in terms of the pion decay constant $F_\pi$~\cite{Adler:1971nq,Terentev:1971cso,Aviv:1971hq},
and, in terms of the KT equations, the $\pi\pi$ rescattering among the final-state pions. 

In the $2\pi$ channel the average is dominated by experiments using the initial-state-radiation (ISR) technique, 
while data from energy-scan experiments~\cite{Akhmetshin:2001ig,Akhmetshin:2003zn,Achasov:2005rg,Achasov:2006vp,Akhmetshin:2006wh,Akhmetshin:2006bx}
are consistent but currently less precise. For the $3\pi$ channel this situation is reversed, with only a single ISR data set, which, in addition, only covers the 
energy region above the $\phi$~\cite{Aubert:2004kj}. Instead, the low-energy region including the $\omega$ and $\phi$ resonances has been most precisely measured by the Novosibirsk experiments
SND~\cite{Achasov:2000am,Achasov:2002ud,Achasov:2003ir,Aulchenko:2015mwt} and CMD-2~\cite{Akhmetshin:1995vz,Akhmetshin:1998se,Akhmetshin:2003zn,Akhmetshin:2006sc}.
For completeness, we will also consider earlier data from DM1~\cite{Cordier:1979qg}, DM2~\cite{Antonelli:1992jx}, and ND~\cite{Dolinsky:1991vq}.
The main part of the paper is then devoted to the fit systematics to the various data sets, as detailed in Sect.~\ref{sec:fits},
before working out the consequences for HVP in Sect.~\ref{sec:amu} and summarizing our findings in Sect.~\ref{sec:summary}.

\section{Dispersive representation of the $\boldsymbol{\gamma^*\to3\pi}$ amplitude}
\label{sec:g3pi}

Neglecting the mass of the electron, the HVP contribution to $(g-2)_\mu$ can be expressed as~\cite{Bouchiat:1961,Brodsky:1967sr}
\beq
	a_\mu = \Big( \frac{\alpha m_\mu}{3\pi} \Big)^2
        \int_{s_\mathrm{thr}}^\infty \diff s \frac{\hat K(s)}{s^2}
        R_\mathrm{had}(s),
\label{eq:amuHVP}
\eeq
with $\alpha=e^2/(4\pi)$, the kernel function
\begin{align}
\hat K(s) &= \frac{3s}{m_\mu^2} \bigg[
		 \frac{x^2}{2} (2-x^2) + \frac{(1+x^2)(1+x)^2}{x^2} \Big( \log(1+x) - x + \frac{x^2}{2} \Big) + \frac{1+x}{1-x} x^2 \log x \bigg] ,\notag \\
		x &= \frac{1-\sigma_\mu(s)}{1+\sigma_\mu(s)},\qquad \sigma_\mu(s)=\sqrt{1-\frac{4m_\mu^2}{s}},
\end{align}
and the hadronic cross section\footnote{Note that $R_\mathrm{had}(s)$ is not exactly the usual $R$ ratio defined as $\sigma(e^+e^-\to\text{hadrons})/\sigma(e^+e^-\to\mu^+\mu^-)$, but
coincides for a tree-level muonic cross section and in the limit $s\gg m_\mu^2$.}
\beq
	\label{eq:Rratio}
	R_\mathrm{had}(s) = \frac{3s}{4\pi\alpha^2}\sigma(e^+e^-\to\mathrm{hadrons}).
\eeq
Since higher-order iterations of HVP do become relevant~\cite{Calmet:1976kd,Kurz:2014wya} (at next-to-leading order, this issue arises, at least in principle, even for HLbL~\cite{Colangelo:2014qya}), 
conventions for the radiative corrections need to be specified. The hadronic cross section is to be understood including final-state radiation (FSR), but with ISR and vacuum polarization (VP)
removed (``bare'' cross section). This issue of radiative corrections is most severe for the $2\pi$ channel, and therein for the ISR data sets, but as demonstrated in~\cite{Campanario:2019mjh}
the corrections are now known sufficiently accurately that they cannot account for the $(g-2)_\mu$ anomaly. 

For the $3\pi$ channel, we have $s_\mathrm{thr}=9\mpi^2$ in~\eqref{eq:amuHVP} and the radiative corrections to the cross section are mainly of conceptual nature. Strictly speaking, a dispersive representation
of the $\gamma^*\to 3\pi$ amplitude is only valid in pure QCD, so that, in principle, all photon contributions including FSR should be removed before the fit and only afterwards added again in a perturbative way. In the case of the $2\pi$ channel~\cite{Colangelo:2018mtw}, this strategy was indeed carried through in the context of a scalar-QED approximation. 
For the $3\pi$ channel, the full HVP contribution is more than an order of magnitude smaller, so that the total size of the $3\pi\gamma$ final state would be naively 
estimated at the level $\lesssim 0.3\times 10^{-10}$, which by itself is borderline relevant at the current level of accuracy. 
However, since FSR is automatically included in the cross sections provided by experiment, the actual effect only concerns a possible distortion of the fit due to subtracting and adding 
the FSR contribution, which will be even smaller and therefore neglected here. In contrast, the VP removal does become relevant at the current level of accuracy, mainly because 
of the resonance enhancement in the vicinity of $\rho$, $\omega$, and $\phi$, which shifts the pole position, see Sect.~\ref{sec:vector_meson_masses}, and modifies the spectral function. 
When provided by experiment, we use the bare cross section directly, otherwise we apply the VP routine from~\cite{Keshavarzi:2018mgv}.
To check the sensitivity to this correction, we also constructed an independent VP function based on the $2\pi$ fit from~\cite{Colangelo:2018mtw} as well as the $3\pi$ cross section 
from the present paper, so that major deviations to the full VP only start in the vicinity of the $\phi$, where the $K\bar K$ channels become relevant. We can therefore check the $3\pi$ contribution
self-consistently up-to-and-including the $\omega$ peak, producing a difference of less than $0.1\times 10^{-10}$ in the HVP integral. Accordingly, 
we conclude that the details of the VP routine lead to a negligible effect as well.

The dispersive representation that we fit to the bare cross section is constructed along the following lines, see~\cite{Hoferichter:2014vra,Hoferichter:2018dmo,Hoferichter:2018kwz} for more details.
First, the cross section is given in terms of the $\gamma^*\to3\pi$ amplitude $\F(s,t,u;q^2)$
according to
\beq
\label{eq:epemcross1}
\sigma_{e^+ e^- \to 3\pi}(q^2) = \alpha^2\int_{s_\text{min}}^{s_\text{max}} \diff s \int_{t_\text{min}}^{t_\text{max}} \diff t \,
\frac{(s-4\mpi^2)\,\lambda(q^2,\mpi^2,s)\sin^2\theta_s}{768 \, \pi \, q^6}  \, |\F(s,t,u;q^2)|^2, 
\eeq
with integration boundaries
\begin{align}
s_\text{min} &= 4 M_\pi^2, \qquad\qquad \,s_\text{max} = \big(\sqrt{q^2}-M_\pi \big)^2, \notag \\ 
t_\text{min/max}&= (E_-^*+E_0^*)^2-\bigg( \sqrt{E_-^{*2}-M_\pi^2} \pm  \sqrt{E_0^{*2}-M_\pi^2} \bigg)^2,
\end{align}
and
\beq
E_-^*=\frac{\sqrt{s}}{2},\qquad E_0^*=\frac{q^2-s-M_\pi^2}{2\sqrt{s}}.
\eeq
The amplitude itself is defined by the matrix element of the electromagnetic current $j_\mu$
\beq
\langle 0|j_\mu(0)|\pi^+(p_+)\pi^-(p_-)\PiO(p_0)\rangle =-\epsilon_{\mu\nu\rho\sigma}\, p_+^{\,\nu}p_-^{\,\rho} p_0^{\,\sigma}\F(s,t,u;q^2),
\eeq
with $q=p_++p_-+p_0$ and kinematics
\begin{align}
\label{kinematics}
s&=(q-p_0)^2,\qquad t=(q-p_+)^2,\qquad u=(q-p_-)^2,\qquad s+t+u=3\mpi^2+q^2,\notag\\
z_s&=\cos\theta_s=\frac{t-u}{\sigma_\pi(s)\lambda^{1/2}(q^2,\mpi^2,s)},\notag\\
\sigma_\pi(s)&=\sqrt{1-\frac{4\mpi^2}{s}},\qquad \lambda(a,b,c)=a^2+b^2+c^2-2(a b + a c +b c).
\end{align}

The constraints from analyticity and unitarity are most conveniently formulated in terms of the partial-wave amplitudes~\cite{Jacob:1959at}
\beq
\F(s,t,u;q^2) =\sum_{l\;{\text{odd}}}f_l(s,q^2)P'_l(z_s),
\eeq
with derivatives of the Legendre polynomials $P'_l(z_s)$. Since higher partial waves are completely irrelevant below the $\rho_3(1690)$ resonance~\cite{Niecknig:2012sj,Hoferichter:2017ftn} 
(see App.~\ref{app:F_waves} for an estimate of the $F$-wave contribution), 
the discontinuity equation reduces to 
\beq
\label{F-dis}
\disc f_1(s, q^2)=2i\,f_1(s,q^2)\,\theta(s-4M_\pi^2)\sin \delta(s)\, e^{-i\delta(s)},
\eeq
where $\delta(s)$ refers to the $\pi\pi$ $P$-wave phase shift. This is where, in a model-independent way, the information about the $\rho(770)$ enters. 
The KT equations define, iteratively, the solution of~\eqref{F-dis} in terms of dispersion integrals involving the Omn\`es function~\cite{Omnes:1958hv}
\beq
\Omega(s) = \exp\bigg\{\frac{s}{\pi}\int_{4M_\pi^2}^{\infty}\diff s'\frac{\delta(s')}{s'(s'-s)}\bigg\},
\eeq
together with deformations of the integration contour necessitated by the decay kinematics.
We solve the KT equations with the $\pi\pi$ phase shift recently extracted from the $e^+e^-\to\pi^+\pi^-$ channel~\cite{Colangelo:2018mtw} and a cutoff parameter $\Lambda_{3\pi}=2.5\GeV$. 
Variations of these input quantities prove irrelevant compared to other sources of systematic uncertainties.

For a given $q^2$, the KT equations determine the $s$-dependence of the partial-wave amplitude $f_1(s,q^2)$, but the overall normalization $a(q^2)$ is not predicted.
At $q^2=0$ it is determined by the low-energy theorem, at $q^2=\Mw^2,\Mphi^2$ it is related to the $\omega,\phi\to3\pi$ decay widths, and in general
it can be extracted from a fit to the $e^+e^-\to 3\pi$ cross section.
We take essentially the same parameterization as in~\cite{Hoferichter:2018dmo,Hoferichter:2018kwz}
\beq
\label{eq:a-par}
a(q^2)=\alpha_A+\frac{q^2}{\pi}\int_{s_\text{thr}}^{\infty} \diff s'\frac{\Im\A (s')}{s'(s'-q^2)}+C_p(q^2),
\eeq
constructed in such a way as to fulfill the low-energy constraint from the chiral anomaly, preserve analyticity of $\F(s,t,u;q^2)$, and be flexible enough to describe the data
up to $1.8\GeV$.  For the exact relation between the partial wave $f_1(s,q^2)$ and its $q^2$-dependent
normalization $a(q^2)$, see~\cite{Hoferichter:2014vra,Hoferichter:2018kwz}.
The significance of the individual terms is as follows: the subtraction constant $\alpha_A$ is determined by the chiral anomaly (corrected by quark-mass renormalization)~\cite{Bijnens:1989ff,Hoferichter:2012pm},
\beq
\label{eq:alphaF3pi}
\alpha_A= \frac{F_{3\pi}}{3}\times 1.066(10),\qquad F_{3\pi}=\frac{1}{4\pi^2 F^3_\pi}.
\eeq
The function $\A$ includes resonant contributions via
\beq
\A(q^2)=\sum_{V}\frac{c_V}{M_V^2-q^2-i\sqrt{q^2}\, \varGamma_V(q^2)},
\eeq
where $V=\omega,\,\phi,\,\omega'(1420),\,\omega''(1650)$. The energy-dependent widths $\varGamma_{\omega/\phi}(q^2)$ of the $\omega/\phi$ mesons
include all the main decay channels, in particular, the phase space for the $3\pi$ decay channels is calculated including $3\pi$ rescattering as well~\cite{Niecknig:2012sj}
and due to the $\omega\to\pi^0\gamma$ channel the integration threshold is $\sthr=\mpii^2$. 
For $\omega$ and $\phi$, the missing channels account for about $2\%$ of the width, which is remedied by a simple rescaling of the partial widths (in~\cite{Hoferichter:2018dmo,Hoferichter:2018kwz} the missing $\omega\to\pi^+\pi^-$ and $\phi\to\eta\gamma$ were also considered explicitly, leading to virtually identical results). 
As before, the parameters for $\omega'$ and $\omega''$  are taken from~\cite{Tanabashi:2018oca}, assuming a $100\%$ branching ratio to $3\pi$, 
but for $\omega$ and $\phi$ we now allow mass and width to vary: with VP removed, noticeable differences to the PDG emerge, see Sect.~\ref{sec:vector_meson_masses}, 
which is expected since the PDG parameters subsume radiative effects. 

Finally, the conformal polynomial in~\eqref{eq:a-par}
\beq
\label{Cp}
C_p(q^2)=\sum_{i=1}^p c_i\big(z(q^2)^i-z(0)^i\big), \qquad z(q^2)=\frac{\sqrt{s_\text{inel}-s_1}-\sqrt{s_\text{inel}-q^2}}{\sqrt{s_\text{inel}-s_1}+\sqrt{s_\text{inel}-q^2}},
\eeq
accounts for non-resonant effects. The inelastic threshold $s_\text{inel}$ is set to $1\GeV^2$ motivated by the nearby $K\bar K$ threshold, the second parameter to $s_1=-1\GeV^2$. 
Further constraints are implemented to remove the $S$-wave cusp in the polynomial and to ensure that the sum rule
\beq
\label{eq:a_SR}
\alpha_A=\frac{1}{\pi}\int_{s_\text{thr}}^{\infty} \diff s'\frac{\Im{a}(s')}{s'}
=\frac{1}{\pi}\int_{s_\text{thr}}^{\infty} \diff s'\frac{\Im\A(s')}{s'}+\frac{1}{\pi}\int_{s_\text{inel}}^{\infty} \diff s'\frac{\Im{C_p}(s')}{s'}
\eeq
is fulfilled exactly. In~\cite{Hoferichter:2018dmo,Hoferichter:2018kwz} we also introduced further parameters to be able to impose a faster asymptotic behavior 
of the imaginary part as required for the dispersive description of the pion transition form factor, but since 
this impaired to some extent the description of the cross section, here, we only consider these additional constraints to estimate systematic uncertainties.

\section{Fits to $\boldsymbol{e^+e^-}$ data}
\label{sec:fits}

\subsection{Data sets and unbiased fitting}
\label{sec:data_sets}

\begin{table}[t]
	\centering
	\renewcommand{\arraystretch}{1.3}
	\small
	\begin{tabular}{l c c c}
	\toprule
	Experiment & Region of $\sqrt{s}$ [GeV] & \# data points & Normalization uncertainty\\
	\midrule
	SND 2002~\cite{Achasov:2000am,Achasov:2002ud} & $[0.98,1.38]$ & $67$ &  $5.0\%$ (data from~\cite{Achasov:2000am})\\
	& & & $5.4\%$ (otherwise)\\ 
	SND 2003~\cite{Achasov:2003ir} & $[0.66,0.97]$ & $49$ & $3.4\%$ for  $\sqrt{s}<0.9\GeV$\\
	&&& $4.5\%$ for  $\sqrt{s}>0.9\GeV$\\
	SND 2015~\cite{Aulchenko:2015mwt} & $[1.05,1.80]$ & $31$ & $3.7\%$\\\midrule
	CMD-2 1995~\cite{Akhmetshin:1995vz} & $[0.99,1.03]$ & $16$ & $4.6\%$\\
	CMD-2 1998~\cite{Akhmetshin:1998se} & $[0.99,1.03]$ & $13$ & $2.3\%$\\
	CMD-2 2004~\cite{Akhmetshin:2003zn} & $[0.76,0.81]$ & $13$ & $1.3\%$\\
	CMD-2 2006~\cite{Akhmetshin:2006sc} & $[0.98,1.06]$ & $54$ & $2.5\%$\\\midrule
	DM1 1980~\cite{Cordier:1979qg} & $[0.75,1.10]$ & $26$ & $3.2\%$\\
	ND 1991~\cite{Dolinsky:1991vq} & $[0.81,1.39]$ & $28$ & $10\%$ for  $\sqrt{s}<1.0\GeV$\\ 
	&&& $20\%$ for  $\sqrt{s}>1.0\GeV$\\
	DM2 1992~\cite{Antonelli:1992jx} & $[1.34,1.80]$ & $10$ & $8.7\%$\\
	BaBar 2004~\cite{Aubert:2004kj} & $[1.06,1.80]$ & $30$ & all systematics\\
	\bottomrule
	\renewcommand{\arraystretch}{1.0}
	\end{tabular}
	\caption{Summary of data sets for $e^+e^-\to3\pi$. For~\cite{Aulchenko:2015mwt,Antonelli:1992jx,Aubert:2004kj} only data points for $\sqrt{s}\leq1.8\GeV$ are included. In the last column we indicate the size of the systematic errors 
	that we interpret as a normalization-type uncertainty and therefore assume to be $100\%$ correlated.}
	\label{tab:data_sets}
\end{table}

We start with a brief summary of the data sets that we will include in our analysis, see Table~\ref{tab:data_sets}. 
For all data sets the statistical errors are given in diagonal form, with the implication that correlations are negligible at least at the quoted level of uncertainty.
In contrast, the treatment of the systematic uncertainties is more ambiguous, since assumptions need to be made on the correlations between data points. 
Some sources of systematic uncertainty are, by definition, $100\%$ correlated, these are normalization uncertainties for instance due to the luminosity measurement and the detection efficiency,
but other systematic effects may well be localized in certain energy regions and therefore should not be considered fully correlated. 
To follow the experimental documentation as closely as possible, we consider a systematic error of normalization-type origin whenever given as a percentage, otherwise,
we treat that uncertainty as a diagonal error. Note that this distinction mainly affects the SND data sets, while for the other energy-scan experiments
all systematic errors are given as a percentage. 
The exception is the ISR data set from BaBar, but~\cite{Aubert:2004kj} states explicitly that the systematic errors for
different mass bins are fully correlated. 

These details are important to monitor a potential bias in the fit. Most importantly, a $\chi^2$-minimization with an empirical full covariance matrix $\mathrm{V}(i,j)$ including a normalization uncertainty,
\beq
\chi^2 = \sum_{i,j} ( f(x_i) - y_i ) \mathrm{V}(i,j)^{-1} ( f(x_j) - y_j ),
\eeq
will converge to a solution that is biased towards a lower value than expected due to the fact that smaller data values are assigned smaller normalization uncertainties. This D'Agostini bias was first observed in~\cite{DAgostini:1993arp}. It becomes increasingly severe for large normalization uncertainties and/or a large number of data points, so precisely
when there is a normalization uncertainty in an experiment that is $100\%$ correlated among all data points. 
In addition, in a global fit of several experiments a bias that may occur in the combination needs to be avoided. 

We follow the iterative fit strategy proposed by the NNPDF collaboration~\cite{Ball:2009qv} to eliminate the bias, which is based on the observation that the normalization uncertainties should be proportional to the true value rather than the measurement. In this manner, the modified iterative covariance matrix is given as 
\beq 
\label{iteration}
\mathrm{V}_{n+1}(i, j)=\mathrm{V}^{\text{stat}}(i, j)+\frac{\mathrm{V}^{\text{syst}}(i, j) }{y_iy_j} f_{n}(x_{i}) f_{n}(x_{j}),
\eeq
where $\mathrm{V}^{\text{stat}}(i, j)$ is the statistical covariance matrix and the systematic covariance matrix $\mathrm{V}^{\text{syst}}(i, j)$ is determined by multiplying the normalization factors with the fit function $f_n(x_i)$ in each iteration step rather than the data. The empirical covariance matrix can be chosen as the initial guess, with expected rapid convergence to the final solution. 

In the fit to the data sets in Table~\ref{tab:data_sets} we only encounter either fully correlated or diagonal errors.  
We follow~\cite{Ball:2009qv} and treat the uncorrelated systematic errors on the same footing as the statistical ones.  
For a single experiment one would therefore expect that the central values obtained in a fit with diagonal errors only should be close to the central values of the full fit,
otherwise, one would need to understand better the role of the correlations. 
In the following, we will thus consider both diagonal and full fits to monitor whether significant differences arise.

\subsection{Fits to SND}

\begin{table}[t]
	\centering
	\footnotesize
	\renewcommand{\arraystretch}{1.3}
	\begin{tabular}{lcccccc}
	\toprule
	& \multicolumn{3}{c}{diagonal} & \multicolumn{3}{c}{full}\\
	$p_\text{conf}$ & $2$ & $3$ & $4$ & $2$ & $3$ & $4$\\
	$\chi^2/\text{dof}$ & $97.6/137$ & $93.5/136$ & $93.2/135$ & $164.9/137$ & $155.4/136$& $152.6/135$\\
	& $=0.71$ & $=0.69$ & $=0.69$ & $=1.20$ & $=1.14$& $=1.13$\\
	$p$-value & $0.996$ & $0.998$ & $0.998$ & $0.052$ & $0.12$& $0.14$\\
	$\Mw \ [\text{MeV}]$ & $782.62(4)$ & $782.62(4)$ & $782.62(4)$ & $782.63(2)$ & $782.63(2)$& $782.63(2)$\\
	$\Gw \ [\text{MeV}]$ & $8.68(6)$ & $8.72(7)$ & $8.73(7)$  & $8.66(3)$ & $8.68(3)$& $8.68(3)$\\
	$\Mphi \ [\text{MeV}]$& $1019.19(4)$ & $1019.18(4)$ & $1019.18(4)$ & $1019.19(2)$ & $1019.19(2)$& $1019.19(2)$\\
	$\Gphi \ [\text{MeV}]$ & $4.16(8)$ & $4.13(8)$ & $4.13(8)$  & $4.17(4)$ & $4.16(4)$& $4.16(4)$\\
	$c_\omega \ [\text{GeV}^{-1}]$ & $2.88(1)$ & $2.89(1)$ & $2.89(1)$ & $2.87(3)$ & $2.88(3)$& $2.90(3)$\\
	$c_\phi \ [\text{GeV}^{-1}]$ & $-0.393(4)$ & $-0.392(4)$ & $-0.392(4)$ & $-0.388(6)$ & $-0.386(6)$& $-0.385(6)$\\
	$c_{\omega'} \ [\text{GeV}^{-1}]$ & $-0.16(4)$ & $-0.08(5)$ & $-0.08(5)$  & $-0.16(3)$ & $-0.06(4)$& $-0.07(5)$\\
	$c_{\omega''} \ [\text{GeV}^{-1}]$ & $-1.59(9)$ & $-1.46(11)$ & $-1.42(14)$ & $-1.62(9)$ & $-1.50(10)$& $-1.42(12)$\\
	$c_1 \ [\text{GeV}^{-3}]$ & $-0.43(11)$ & $-0.33(13)$ & $-0.32(13)$ & $-0.37(11)$ & $-0.18(12)$& $-0.06(15)$\\
	$c_2 \ [\text{GeV}^{-3}]$ & $-1.35(5)$ & $-1.44(7)$ & $-1.49(12)$ & $-1.30(5)$ & $-1.42(6)$& $-1.58(12)$\\
	$c_3 \ [\text{GeV}^{-3}]$ & --- & $-0.45(9)$ & $-0.41(12)$ & --- & $-0.48(8)$& $-0.41(10)$\\
	$c_4 \ [\text{GeV}^{-3}]$ & --- & --- & $1.40(10)$ & --- & --- & $1.52(10)$ \\
	$10^{10}\times a_\mu^{3\pi}|_{\leq 1.8\GeV}$ & $47.28(25)$ & $47.31(25)$ & $47.34(25)$ & $46.74(92)$ & $46.97(93)$ & $47.53(1.00)$\\ 
	\bottomrule
	\renewcommand{\arraystretch}{1.0}
	\end{tabular}
	\caption{Fits to the combination of SND data sets~\cite{Achasov:2000am,Achasov:2002ud,Achasov:2003ir,Aulchenko:2015mwt}, for diagonal errors and full covariance matrices. $p_\text{conf}$ denotes the number of free parameters in the conformal polynomial. All errors refer to fit uncertainties only.}
	\label{tab:fits_SND}
\end{table}

As the first set of fits we consider the SND data sets~\cite{Achasov:2000am,Achasov:2002ud,Achasov:2003ir,Aulchenko:2015mwt}. The results are summarized in Table~\ref{tab:fits_SND}, both for
diagonal errors only and including correlations as described in the previous section. In each case we consider variants of the fits with $p_\text{conf}=2\ldots 4$ free parameters
in the conformal polynomial and at this stage display only the fit uncertainties, with systematic uncertainties of the dispersive representation to be added later.

The results in Table~\ref{tab:fits_SND} show that the main effect of the correlations is an increase in the uncertainty, within the fit statistics the central values agree with the 
diagonal fit. However, we also note that the description of the data becomes worse, which can be remedied to some extent by increasing $p_\text{conf}$. 
While the diagonal fit proves very stable to variations of $p_\text{conf}$, we observe that when including the correlations the central value increases with $p_\text{conf}$,
balancing the reduction in the central value compared to the diagonal fit in the variant with $p_\text{conf}=2$, the smallest for which a reasonable fit can be obtained.

We note that the treatment of the systematic uncertainties, closely following experiment as specified in Sect.~\ref{sec:data_sets}, is critical to obtain consistent fits. 
If all systematic uncertainties were assumed to be fully correlated, the fit iteration would not even converge or, when restricted to a subset of the data,
lead to a significant downward bias.

\subsection{Fits to CMD-2 and BaBar}

\begin{table}[t]
	\centering
	\footnotesize
	\renewcommand{\arraystretch}{1.3}
	\begin{tabular}{lcccccc}
	\toprule
	& \multicolumn{3}{c}{diagonal} & \multicolumn{3}{c}{full}\\
	$p_\text{conf}$ & $2$ & $3$ & $4$ & $2$ & $3$ & $4$\\
	$\chi^2/\text{dof}$ & $83.9/62$ & $83.5/61$ & $77.4/60$ & $91.9/62$ & $91.6/61$& $84.3/60$\\
	& $=1.35$ & $=1.37$ & $=1.29$ & $=1.48$ & $=1.50$& $=1.41$\\
	$p$-value & $0.03$ & $0.03$ & $0.06$ & $0.008$ & $0.007$& $0.02$\\
	$\Mw \ [\text{MeV}]$ & $782.49(10)$ & $782.49(10)$ & $782.50(10)$ & $782.49(9)$ & $782.49(9)$& $782.50(9)$\\
	$\Gw \ [\text{MeV}]$ & $9.11(17)$ & $9.13(16)$ & $8.99(16)$  & $9.11(15)$ & $9.13(15)$& $9.00(15)$\\
	$\Mphi \ [\text{MeV}]$& $1019.25(4)$ & $1019.25(4)$ & $1019.22(4)$ & $1019.28(4)$ & $1019.27(4)$& $1019.25(4)$\\
	$\Gphi \ [\text{MeV}]$ & $4.46(11)$ & $4.45(11)$ & $4.45(11)$  & $4.46(10)$ & $4.46(10)$& $4.46(10)$\\
	$c_\omega \ [\text{GeV}^{-1}]$ & $2.91(4)$ & $2.92(4)$ & $2.88(4)$ & $2.91(4)$ & $2.92(4)$& $2.88(4)$\\
	$c_\phi \ [\text{GeV}^{-1}]$ & $-0.406(8)$ & $-0.406(8)$ & $-0.407(8)$ & $-0.405(8)$ & $-0.404(8)$& $-0.405(8)$\\
	$c_{\omega'} \ [\text{GeV}^{-1}]$ & $-0.25(11)$ & $-0.21(13)$ & $-0.19(14)$  & $-0.24(11)$ & $-0.21(12)$& $-0.18(13)$\\
	$c_{\omega''} \ [\text{GeV}^{-1}]$ & $-2.03(32)$ & $-1.97(31)$ & $-2.69(37)$ & $-2.01(31)$ & $-1.98(30)$& $-2.73(35)$\\
	$c_1 \ [\text{GeV}^{-3}]$ & $0.12(43)$ & $0.22(42)$ & $0.20(30)$ & $0.07(43)$ & $0.17(43)$& $0.10(29)$\\
	$c_2 \ [\text{GeV}^{-3}]$ & $-1.14(12)$ & $-1.19(14)$ & $-0.31(40)$ & $-1.16(11)$ & $-1.19(13)$& $-0.24(39)$\\
	$c_3 \ [\text{GeV}^{-3}]$ & --- & $-0.84(28)$ & $-1.51(34)$ & --- & $-0.84(28)$& $-1.52(32)$\\
	$c_4 \ [\text{GeV}^{-3}]$ & --- & --- & $1.20(22)$ & --- & --- & $1.19(21)$ \\
	$10^{10}\times a_\mu^{3\pi}|_{\leq 1.8\GeV}$ & $46.17(56)$ & $46.19(55)$ & $45.80(57)$ & $46.23(74)$ & $46.27(74)$ & $45.83(75)$\\ 
	\bottomrule
	\renewcommand{\arraystretch}{1.0}
	\end{tabular}
	\caption{Fits to the combination of the CMD-2 data sets~\cite{Akhmetshin:1995vz,Akhmetshin:1998se,Akhmetshin:2003zn} and BaBar~\cite{Aubert:2004kj}.}
	\label{tab:fits_CMD2_9598}
\end{table}

The CMD-2 data sets mainly cover the resonance regions, with~\cite{Akhmetshin:2003zn} scattered around the $\omega$ peak and~\cite{Akhmetshin:1995vz,Akhmetshin:1998se,Akhmetshin:2006sc} 
around the $\phi$. To be able to perform fits to the whole energy region up to $1.8\GeV$ and thus facilitate the comparison to the SND fits we combine the CMD-2 data with the BaBar
data set~\cite{Aubert:2004kj}, which starts directly above the $\phi$ and covers the remainder.  

We do not find acceptable fits for the naive combination of all these data sets. To isolate the reason we perform two separate fits, first, to~\cite{Akhmetshin:1995vz,Akhmetshin:1998se,Akhmetshin:2003zn} and BaBar~\cite{Aubert:2004kj}, as given in Table~\ref{tab:fits_CMD2_9598}, as well as~\cite{Akhmetshin:2003zn,Akhmetshin:2006sc} and BaBar~\cite{Aubert:2004kj}, see Table~\ref{tab:fits_CMD2_2006}. This strategy is motivated by the suspicion that inconsistencies among the CMD-2 data sets arise in the vicinity of the $\phi$, which the separate consideration
of the data sets covering this region should be able to corroborate.

\begin{table}[t]
	\centering
	\footnotesize
	\renewcommand{\arraystretch}{1.3}
	\begin{tabular}{lcccccc}
	\toprule
	& \multicolumn{3}{c}{diagonal} & \multicolumn{3}{c}{full}\\
	$p_\text{conf}$ & $2$ & $3$ & $4$ & $2$ & $3$ & $4$\\
	$\chi^2/\text{dof}$ & $199.4/87$ & $164.8/86$ & $142.0/85$ & $213.3/87$ & $185.0/86$& $156.4/85$\\
	& $=2.29$ & $=1.91$ & $=1.67$ & $=2.45$ & $=2.15$& $=1.84$\\
	$p$-value & $8\times 10^{-11} $ & $7\times 10^{-7}$ & $1\times 10^{-4}$ & $1\times 10^{-12}$ & $3\times 10^{-9}$& $4\times 10^{-6}$\\
	$\Mw \ [\text{MeV}]$ & $782.53(10)$ & $782.54(10)$ & $782.56(10)$ & $782.53(10)$ & $782.54(9)$& $782.56(10)$\\
	$\Gw \ [\text{MeV}]$ & $8.74(12)$ & $8.74(12)$ & $8.48(13)$  & $8.76(13)$ & $8.80(12)$& $8.56(13)$\\
	$\Mphi \ [\text{MeV}]$& $1019.11(2)$ & $1019.09(2)$ & $1019.07(2)$ & $1019.10(2)$ & $1019.08(2)$& $1019.07(2)$\\
	$\Gphi \ [\text{MeV}]$ & $4.40(6)$ & $4.34(6)$ & $4.34(6)$  & $4.35(5)$ & $4.30(5)$& $4.29(5)$\\
	$c_\omega \ [\text{GeV}^{-1}]$ & $2.81(3)$ & $2.81(3)$ & $2.74(3)$ & $2.80(3)$ & $2.81(3)$& $2.74(3)$\\
	$c_\phi \ [\text{GeV}^{-1}]$ & $-0.399(4)$ & $-0.396(4)$ & $-0.396(4)$ & $-0.408(6)$ & $-0.401(6)$& $-0.401(6)$\\
	$c_{\omega'} \ [\text{GeV}^{-1}]$ & $-0.79(8)$ & $-0.19(12)$ & $-0.09(13)$  & $-0.63(9)$ & $-0.18(11)$& $-0.08(12)$\\
	$c_{\omega''} \ [\text{GeV}^{-1}]$ & $-3.08(19)$ & $-1.92(26)$ & $-2.67(29)$ & $-2.76(23)$ & $-1.90(25)$& $-2.73(28)$\\
	$c_1 \ [\text{GeV}^{-3}]$ & $1.77(29)$ & $1.43(27)$ & $0.54(30)$ & $1.38(32)$ & $1.30(27)$& $0.37(29)$\\
	$c_2 \ [\text{GeV}^{-3}]$ & $-0.27(11)$ & $-0.97(14)$ & $0.25(32)$ & $-0.47(11)$ & $-0.99(13)$& $0.35(32)$\\
	$c_3 \ [\text{GeV}^{-3}]$ & --- & $-0.60(23)$ & $-1.35(27)$ & --- & $-0.60(22)$& $-1.40(25)$\\
	$c_4 \ [\text{GeV}^{-3}]$ & --- & --- & $1.34(21)$ & --- & --- & $1.28(20)$ \\
	$10^{10}\times a_\mu^{3\pi}|_{\leq 1.8\GeV}$ & $44.36(48)$ & $44.40(48)$ & $43.87(49)$ & $44.10(66)$ & $44.32(66)$ & $43.74(66)$\\ 
	\bottomrule
	\renewcommand{\arraystretch}{1.0}
	\end{tabular}
	\caption{Fits to the combination of the CMD-2 data sets~\cite{Akhmetshin:2003zn,Akhmetshin:2006sc} and BaBar~\cite{Aubert:2004kj}.}
	\label{tab:fits_CMD2_2006}
\end{table}

In all cases we see that the diagonal and full fits are well compatible, so that the treatment of correlations becomes less of a concern than for the SND fits. 
However, we find that the fit quality is quite poor: while for the~\cite{Akhmetshin:1995vz,Akhmetshin:1998se} $\phi$ data sets the fits with $p_\text{conf}=4$ might still be considered acceptable,
this is certainly not the case for~\cite{Akhmetshin:2006sc}, even though also in this case higher orders in the conformal expansion do yield some improvement 
of the $\chi^2$.
The most relevant discrepancies in the fit results concern the $\omega$ coupling $c_\omega$, which is significantly smaller in Table~\ref{tab:fits_CMD2_2006} despite being based on the same 
data set in the $\omega$ region, leading to the overall much lower HVP integral, as well as the $\phi$ mass. The fits in Table~\ref{tab:fits_CMD2_9598} prefer a value
around $\Mphi=1019.25(4)\MeV$, while the fits in Table~\ref{tab:fits_CMD2_2006} point to $\Mphi=1019.09(3)\MeV$, suggesting that inconsistencies in the $\phi$ region are 
compensated elsewhere in the fit, thus the change in $c_\omega$.

From the mass shifts discussed in App.~\ref{app:mass_shifts}, together with the PDG $\phi$ mass, we would expect a fit value $\Mphi=1019.20\MeV$, 
in perfect agreement with Table~\ref{tab:fits_SND}, largely consistent with Table~\ref{tab:fits_CMD2_9598}, but clearly at odds with Table~\ref{tab:fits_CMD2_2006}.
Since within uncertainties the $\omega$ masses are consistent among the three fits, this suggests as a remedy to include energy-calibration uncertainties
in the context of~\cite{Akhmetshin:2006sc}, in analogy to the energy rescalings found necessary in the case of the $2\pi$ channel~\cite{Colangelo:2018mtw}.
In fact, \cite{Akhmetshin:2006sc} includes three different scans, and separate fits to each of them reveal that the first two yield $\phi$ masses in the expected range, while the third one
differs, leading to the lower mass in Table~\ref{tab:fits_CMD2_2006}. Accordingly, we apply a rescaling
\beq
\sqrt{s}\to\sqrt{s} +\xi (\sqrt{s}-3\mpi)
\eeq
to the data of the third scan only. The fit prefers a rescaling around $\xi\sim 10^{-4}$, well in line with potential uncertainties of the energy calibration. 
Including $\xi$ as an additional parameter in the fit indeed leads to a mild improvement in the $\chi^2$.

\begin{table}[t]
	\centering
	\footnotesize
	\renewcommand{\arraystretch}{1.3}
	\begin{tabular}{lcccccc}
	\toprule
	& \multicolumn{3}{c}{diagonal} & \multicolumn{3}{c}{full}\\
	$p_\text{conf}$ & $2$ & $3$ & $4$ & $2$ & $3$ & $4$\\
	$\chi^2/\text{dof}$ & $117.9/76$ & $111.8/75$ & $97.7/74$ & $133.9/76$ & $129.1/75$& $112.2/74$\\
	& $=1.55$ & $=1.49$ & $=1.32$ & $=1.76$ & $=1.72$& $=1.52$\\
	$p$-value & $0.001$ & $0.004$ & $0.03$ & $5\times 10^{-5}$ & $1\times 10^{-4}$& $0.003$\\
	$\Mw \ [\text{MeV}]$ & $782.50(10)$ & $782.50(10)$ & $782.52(10)$ & $782.51(10)$ & $782.51(9)$& $782.53(10)$\\
	$\Gw \ [\text{MeV}]$ & $8.95(16)$ & $9.01(14)$ & $8.79(15)$  & $8.96(15)$ & $9.01(14)$& $8.82(14)$\\
	$\Mphi \ [\text{MeV}]$& $1019.19(3)$ & $1019.18(3)$ & $1019.15(3)$ & $1019.17(2)$ & $1019.16(2)$& $1019.14(3)$\\
	$\Gphi \ [\text{MeV}]$ & $4.34(6)$ & $4.31(6)$ & $4.30(6)$  & $4.31(5)$ & $4.29(5)$& $4.28(5)$\\
	$c_\omega \ [\text{GeV}^{-1}]$ & $2.87(4)$ & $2.88(3)$ & $2.83(3)$ & $2.86(4)$ & $2.88(3)$& $2.82(4)$\\
	$c_\phi \ [\text{GeV}^{-1}]$ & $-0.393(4)$ & $-0.392(4)$ & $-0.392(4)$ & $-0.397(6)$ & $-0.395(6)$& $-0.394(6)$\\
	$c_{\omega'} \ [\text{GeV}^{-1}]$ & $-0.37(12)$ & $-0.19(13)$ & $-0.13(14)$  & $-0.34(11)$ & $-0.18(12)$& $-0.12(12)$\\
	$c_{\omega''} \ [\text{GeV}^{-1}]$ & $-2.20(35)$ & $-1.94(29)$ & $-2.73(32)$ & $-2.17(34)$ & $-1.94(27)$& $-2.80(31)$\\
	$c_1 \ [\text{GeV}^{-3}]$ & $0.41(48)$ & $0.66(36)$ & $0.28(29)$ & $0.39(47)$ & $0.60(35)$& $0.15(28)$\\
	$c_2 \ [\text{GeV}^{-3}]$ & $-0.95(13)$ & $-1.14(14)$ & $-0.03(35)$ & $-0.98(12)$ & $-1.14(14)$& $0.09(35)$\\
	$c_3 \ [\text{GeV}^{-3}]$ & --- & $-0.75(25)$ & $-1.51(30)$ & --- & $-0.76(25)$& $-1.54(28)$\\
	$c_4 \ [\text{GeV}^{-3}]$ & --- & --- & $1.26(22)$ & --- & --- & $1.23(21)$ \\
	$10^4\times \xi$ & $1.4(7)$ & $1.4(7)$ & $1.4(7)$ & $1.1(6)$ & $1.0(6)$ & $1.0(6)$ \\
	$10^{10}\times a_\mu^{3\pi}|_{\leq 1.8\GeV}$ & $45.35(54)$ & $45.42(51)$ & $44.89(52)$ & $45.26(72)$ & $45.38(70)$ & $44.87(70)$\\ 
	\bottomrule
	\renewcommand{\arraystretch}{1.0}
	\end{tabular}
	\caption{Fits to the combination of the CMD-2 data sets~\cite{Akhmetshin:2003zn,Akhmetshin:2006sc} and BaBar~\cite{Aubert:2004kj}, with modifications
	to~\cite{Akhmetshin:2006sc} as described in the main text.}
	\label{tab:fits_CMD2_2006_mod}
\end{table}

However, removing this tension in $\Mphi$ by no means renders the resulting fits statistically acceptable. Inspection of the contribution to the $\chi^2$ from each data point 
shows that a by far disproportionate amount originates from the last few points of each scan of~\cite{Akhmetshin:2006sc} for which the cross section drops below $5\,\text{nb}$. 
In the end, we have to conclude that these points cannot be described in a statistically acceptable way with our dispersive representation. 
To demonstrate the huge impact on the fit, Table~\ref{tab:fits_CMD2_2006_mod} gives the results when these critical points are removed.
The fit is clearly still not perfect, but at least comparable in quality to Table~\ref{tab:fits_CMD2_9598}. 
Accordingly, we believe that there is reason to suspect some additional systematic uncertainty in the off-peak cross sections from~\cite{Akhmetshin:2006sc}
and therefore will only consider the reduced data set as in Table~\ref{tab:fits_CMD2_2006_mod} in the following (denoted by CMD-2$'$).

\subsection{Combined fits}
\label{sec:combined}

\begin{table}[t]
	\centering
	\footnotesize
	\renewcommand{\arraystretch}{1.3}
	\begin{tabular}{lcccccc}
	\toprule
	& \multicolumn{3}{c}{diagonal} & \multicolumn{3}{c}{full}\\
	$p_\text{conf}$ & $2$ & $3$ & $4$ & $2$ & $3$ & $4$\\
	$\chi^2/\text{dof}$ & $361.3/306$ & $354.6/305$ & $354.0/304$ & $443.7/306$ & $430.8/305$& $430.7/304$\\
	& $=1.18$ & $=1.16$ & $=1.16$ & $=1.45$ & $=1.41$& $=1.42$\\
	$p$-value & $0.02$ & $0.03$ & $0.03$ & $4\times 10^{-7}$ & $3\times 10^{-6}$& $2\times 10^{-6}$\\
	$\Mw \ [\text{MeV}]$ & $782.60(4)$ & $782.60(4)$ & $782.60(4)$ & $782.63(2)$ & $782.63(2)$& $782.63(2)$\\
	$\Gw \ [\text{MeV}]$ & $8.75(6)$ & $8.79(6)$ & $8.77(6)$  & $8.69(3)$ & $8.71(3)$& $8.71(3)$\\
	$\Mphi \ [\text{MeV}]$& $1019.23(2)$ & $1019.22(2)$ & $1019.22(2)$ & $1019.20(1)$ & $1019.20(1)$& $1019.20(1)$\\
	$\Gphi \ [\text{MeV}]$ & $4.34(4)$ & $4.32(4)$ & $4.32(4)$  & $4.24(3)$ & $4.23(3)$& $4.23(3)$\\
	$c_\omega \ [\text{GeV}^{-1}]$ & $2.87(1)$ & $2.89(1)$ & $2.88(1)$ & $2.85(2)$ & $2.86(2)$& $2.86(2)$\\
	$c_\phi \ [\text{GeV}^{-1}]$ & $-0.395(3)$ & $-0.394(3)$ & $-0.394(3)$ & $-0.388(3)$ & $-0.386(3)$& $-0.386(3)$\\
	$c_{\omega'} \ [\text{GeV}^{-1}]$ & $-0.18(3)$ & $-0.09(5)$ & $-0.08(5)$  & $-0.17(3)$ & $-0.07(4)$& $-0.06(4)$\\
	$c_{\omega''} \ [\text{GeV}^{-1}]$ & $-1.65(8)$ & $-1.52(10)$ & $-1.55(10)$ & $-1.65(8)$ & $-1.52(8)$& $-1.53(10)$\\
	$c_1 \ [\text{GeV}^{-3}]$ & $-0.35(10)$ & $-0.22(11)$ & $-0.24(11)$ & $-0.31(10)$ & $-0.12(11)$& $-0.14(12)$\\
	$c_2 \ [\text{GeV}^{-3}]$ & $-1.28(4)$ & $-1.39(6)$ & $-1.33(9)$ & $-1.24(4)$ & $-1.36(5)$& $-1.34(9)$\\
	$c_3 \ [\text{GeV}^{-3}]$ & --- & $-0.48(8)$ & $-0.51(9)$ & --- & $-0.47(7)$& $-0.48(8)$\\
	$c_4 \ [\text{GeV}^{-3}]$ & --- & --- & $1.39(9)$ & --- & --- & $1.41(9)$ \\
	$10^4\times \xi$ & $1.9(7)$ & $1.8(7)$ & $1.8(7)$ & $1.3(5)$ & $1.3(5)$ & $1.3(5)$ \\
	$10^{10}\times a_\mu^{3\pi}|_{\leq 1.8\GeV}$ & $46.65(21)$ & $46.70(21)$ & $46.67(22)$ & $45.87(47)$ & $46.16(47)$ & $46.10(50)$\\ 
	\bottomrule
	\renewcommand{\arraystretch}{1.0}
	\end{tabular}
	\caption{Fits to the combination of SND~\cite{Achasov:2000am,Achasov:2002ud,Achasov:2003ir,Aulchenko:2015mwt}, 
	CMD-2$'$~\cite{Akhmetshin:1995vz,Akhmetshin:1998se,Akhmetshin:2003zn,Akhmetshin:2006sc}, BaBar~\cite{Aubert:2004kj}, DM1~\cite{Cordier:1979qg}, and ND~\cite{Dolinsky:1991vq}.}
	\label{tab:fits_combination}
\end{table}

Our final preferred fit is shown in Table~\ref{tab:fits_combination}, including all data sets listed in Table~\ref{tab:data_sets} except for the DM2 data~\cite{Antonelli:1992jx},
which disagree with both the BaBar~\cite{Aubert:2004kj} and the SND~\cite{Aulchenko:2015mwt} data especially in the vicinity of the $\omega''(1650)$. 
We also considered fits dropping the CMD-2 data set~\cite{Akhmetshin:2006sc} altogether, see Table~\ref{tab:fits_combination_wo_CMD2_2006}, but
the overall effect is relatively minor, depending on the fit variant at most $0.2\times 10^{-10}$ in the final $(g-2)_\mu$ integral.

\begin{figure}[t]
	\centering
	\includegraphics[width=\linewidth,clip]{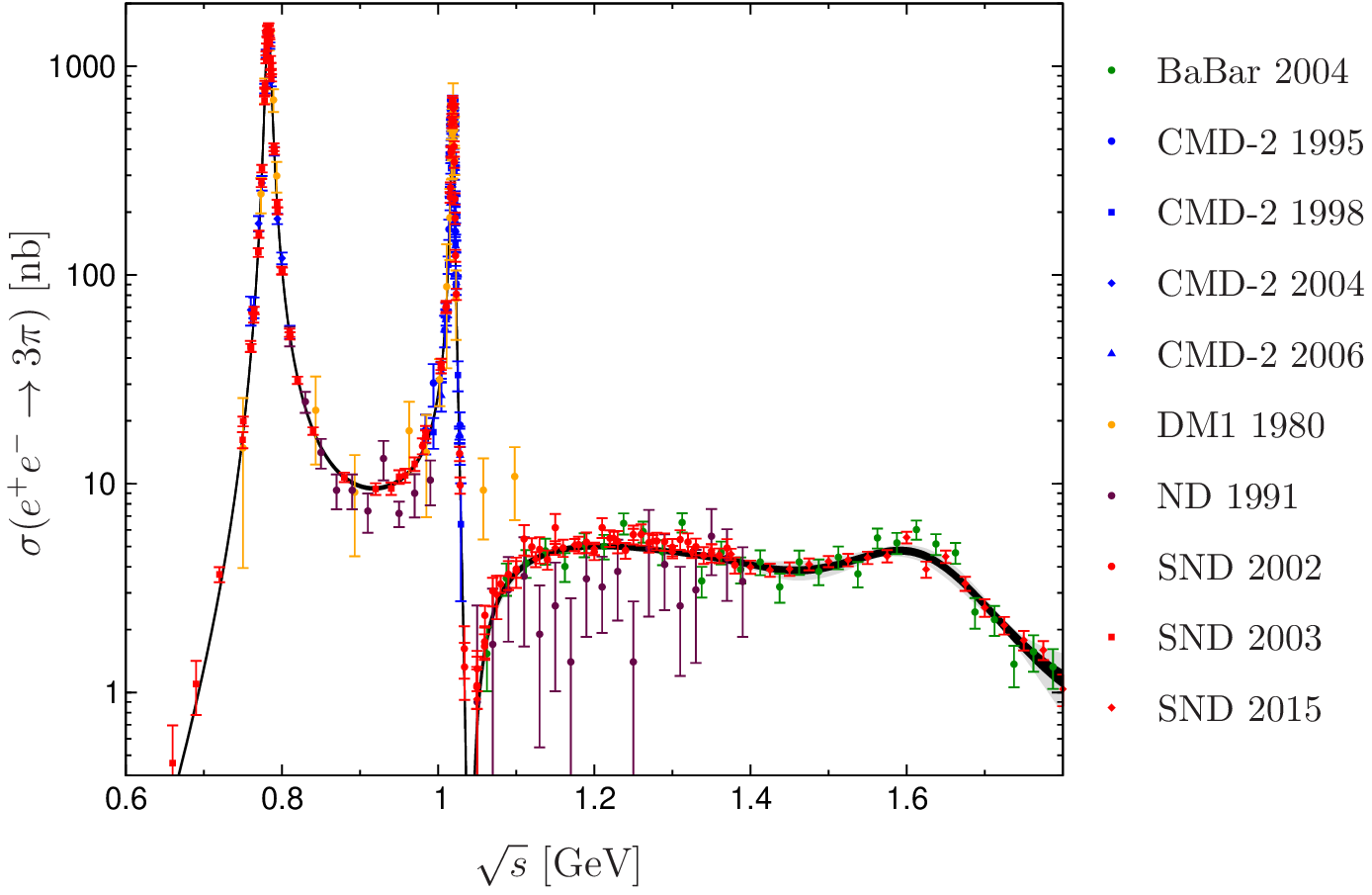}\\[4mm]
	\includegraphics[width=0.49\linewidth,clip]{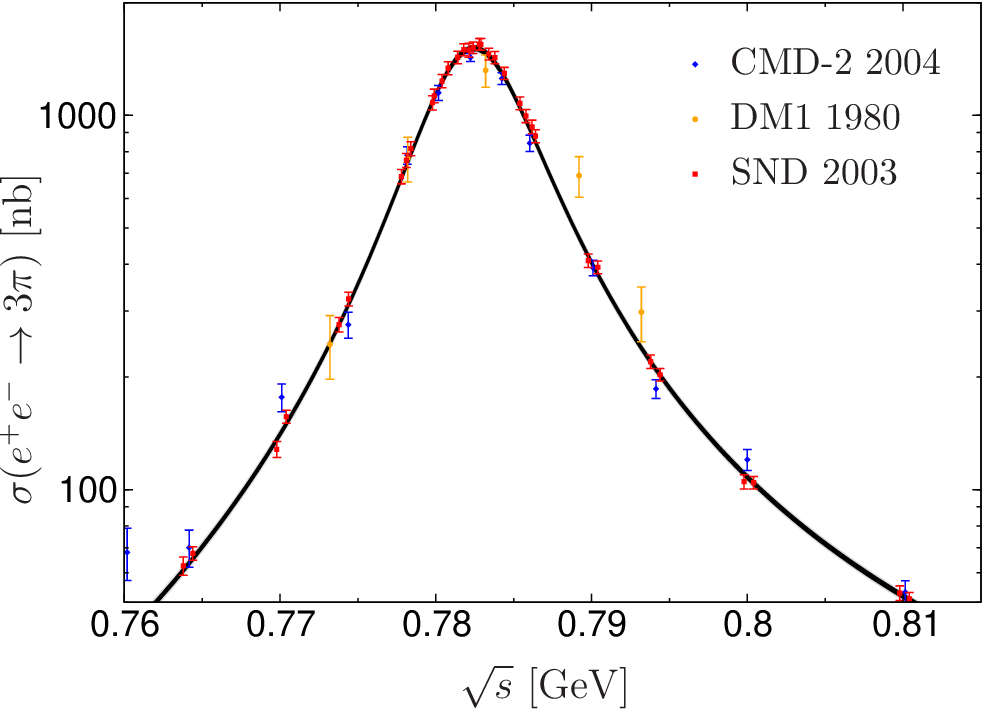}
	\includegraphics[width=0.49\linewidth,clip]{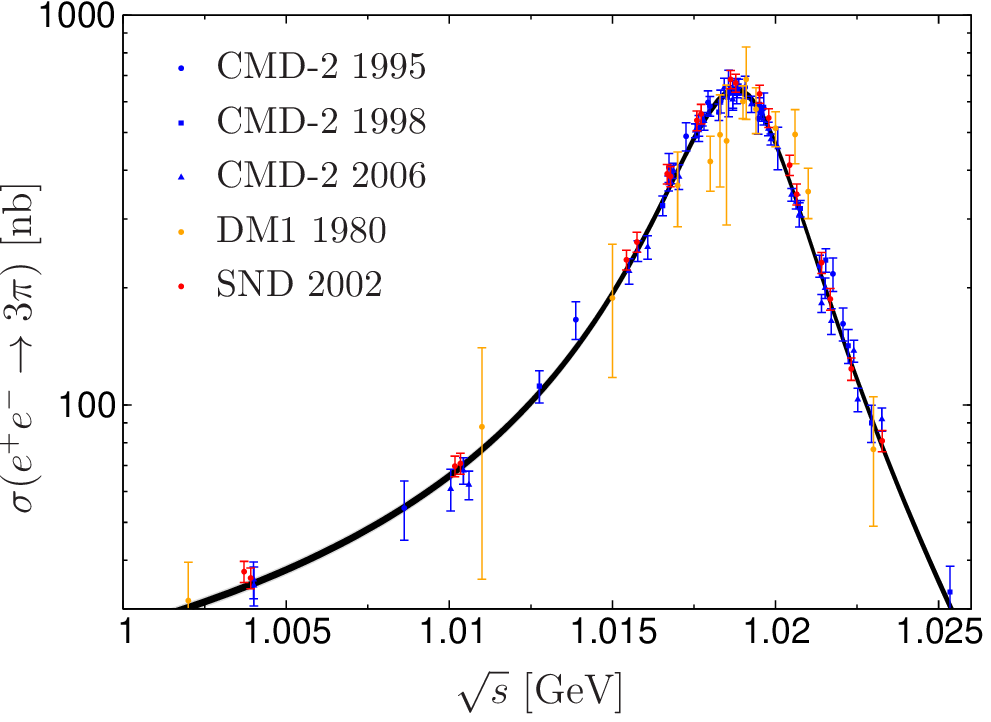}
	\caption{Fit to the $e^+e^-\to3\pi$ data sets as listed in Table~\ref{tab:data_sets} (with VP removed everywhere). The black band includes the fit uncertainties only, while the gray band represents the total uncertainty, including the systematics of the dispersive representation.
	For most energies the two uncertainties are of similar size, so that the difference is hardly visible on the logarithmic scale.
	}
	\label{fig:cross_section}
\end{figure}

In all cases the $\chi^2$ is significantly worse than in the separate fits discussed in the previous sections. Since there are now several experiments covering the same energy region,
this is an indication of the degree of consistency among the various data sets. To account for these inconsistencies we follow the PDG prescription~\cite{Tanabashi:2018oca} 
and inflate the fit errors by the scale factor
\beq
S=\sqrt{\chi^2/\text{dof}},
\eeq
which increases uncertainties by about $20\%$ compared to the fit errors given in Table~\ref{tab:fits_combination}.
The systematic uncertainties are dominated by the degree of the conformal polynomial, while the uncertainties from $\pi\pi$ phase shifts and cutoff parameters are negligible in comparison.
We adopt the full results for $p_\text{conf}=3$ as our central value (as this gives the best fit), but keep the maximum differences to $p_\text{conf}=2,4$ as a source of systematic uncertainty.
In addition, we perform fits in which the imaginary part of the conformal polynomial in~\eqref{Cp} is constrained to behave as $q^{-3}$ asymptotically, and 
include the observed variation as another source of systematics, see Table~\ref{tab:fits_combination_asym} for this last set of fits. As alluded to earlier, the fit quality 
deteriorates when imposing this additional constraint on the conformal polynomial, and therefore the full variation over all fit variants would likely be an overestimate of the systematic uncertainty.
To gauge the impact, we take the average change for $p_\text{conf}=2,3,4$ separately, 
and add the result in quadrature to the systematic error from the variation in $p_\text{conf}$.\footnote{Note that the fits with $p_\text{conf}=4$ from Table~\ref{tab:fits_combination_asym} already display signs of numerical instabilities, with large shifts in the fit parameters compared to $p_\text{conf}=3$ and sizable cancellations among the terms 
in the conformal polynomial. We still include this fit in the estimate of the systematic uncertainties, otherwise, the systematic errors of the final results 
given in Sects.~\ref{sec:vector_meson_masses} and~\ref{sec:amu} would decrease slightly.}
The final fit is illustrated in Fig.~\ref{fig:cross_section}.

\begin{table}[t]
	\centering
	\footnotesize
	\renewcommand{\arraystretch}{1.3}
	\begin{tabular}{lcccccc}
	\toprule
	& \multicolumn{3}{c}{diagonal} & \multicolumn{3}{c}{full}\\
	$p_\text{conf}$ & $2$ & $3$ & $4$ & $2$ & $3$ & $4$\\
	$\chi^2/\text{dof}$ & $286.5/263$ & $283.3/262$ & $283.3/261$ & $368.0/263$ & $358.9/262$& $358.6/261$\\
	& $=1.09$ & $=1.08$ & $=1.09$ & $=1.40$ & $=1.37$& $=1.37$\\
	$p$-value & $0.15$ & $0.19$ & $0.16$ & $2\times 10^{-5}$ & $6\times 10^{-5}$& $6\times 10^{-5}$\\
	$\Mw \ [\text{MeV}]$ & $782.60(4)$ & $782.60(4)$ & $782.60(4)$ & $782.63(2)$ & $782.63(2)$& $782.63(2)$\\
	$\Gw \ [\text{MeV}]$ & $8.77(6)$ & $8.80(6)$ & $8.80(6)$  & $8.70(3)$ & $8.71(3)$& $8.71(3)$\\
	$\Mphi \ [\text{MeV}]$& $1019.24(3)$ & $1019.23(3)$ & $1019.23(3)$ & $1019.22(2)$ & $1019.21(2)$& $1019.21(2)$\\
	$\Gphi \ [\text{MeV}]$ & $4.28(6)$ & $4.26(6)$ & $4.26(6)$  & $4.21(4)$ & $4.20(4)$& $4.20(4)$\\
	$c_\omega \ [\text{GeV}^{-1}]$ & $2.88(1)$ & $2.89(1)$ & $2.89(1)$ & $2.85(2)$ & $2.87(2)$& $2.87(2)$\\
	$c_\phi \ [\text{GeV}^{-1}]$ & $-0.395(4)$ & $-0.394(4)$ & $-0.394(4)$ & $-0.385(4)$ & $-0.384(4)$& $-0.383(4)$\\
	$c_{\omega'} \ [\text{GeV}^{-1}]$ & $-0.17(3)$ & $-0.10(5)$ & $-0.10(5)$  & $-0.17(3)$ & $-0.08(4)$& $-0.08(4)$\\
	$c_{\omega''} \ [\text{GeV}^{-1}]$ & $-1.65(8)$ & $-1.55(9)$ & $-1.56(11)$ & $-1.67(8)$ & $-1.55(8)$& $-1.53(10)$\\
	$c_1 \ [\text{GeV}^{-3}]$ & $-0.36(10)$ & $-0.27(11)$ & $-0.27(12)$ & $-0.30(10)$ & $-0.14(11)$& $-0.11(12)$\\
	$c_2 \ [\text{GeV}^{-3}]$ & $-1.30(4)$ & $-1.38(6)$ & $-1.37(10)$ & $-1.25(4)$ & $-1.36(5)$& $-1.39(10)$\\
	$c_3 \ [\text{GeV}^{-3}]$ & --- & $-0.51(8)$ & $-0.51(9)$ & --- & $-0.51(7)$& $-0.49(8)$\\
	$c_4 \ [\text{GeV}^{-3}]$ & --- & --- & $1.34(9)$ & --- & --- & $1.40(9)$ \\
	$10^{10}\times a_\mu^{3\pi}|_{\leq 1.8\GeV}$ & $46.81(22)$ & $46.84(22)$ & $46.84(22)$ & $46.02(50)$ & $46.21(50)$ & $46.29(53)$\\ 
	\bottomrule
	\renewcommand{\arraystretch}{1.0}
	\end{tabular}
	\caption{Fits to the combination of SND~\cite{Achasov:2000am,Achasov:2002ud,Achasov:2003ir,Aulchenko:2015mwt}, 
	CMD-2~\cite{Akhmetshin:1995vz,Akhmetshin:1998se,Akhmetshin:2003zn}, BaBar~\cite{Aubert:2004kj}, DM1~\cite{Cordier:1979qg}, and ND~\cite{Dolinsky:1991vq}.}
	\label{tab:fits_combination_wo_CMD2_2006}
\end{table}

\begin{table}[t]
	\centering
	\footnotesize
	\renewcommand{\arraystretch}{1.3}
	\begin{tabular}{lcccccc}
	\toprule
	& \multicolumn{3}{c}{diagonal} & \multicolumn{3}{c}{full}\\
	$p_\text{conf}$ & $2$ & $3$ & $4$ & $2$ & $3$ & $4$\\
	$\chi^2/\text{dof}$ & $382.8/306$ & $382.7/305$ & $353.2/304$ & $469.5/306$ & $469.5/305$& $432.3/304$\\
	& $=1.25$ & $=1.25$ & $=1.16$ & $=1.53$ & $=1.54$& $=1.42$\\
	$p$-value & $0.002$ & $0.002$ & $0.03$ & $5\times 10^{-9}$ & $4\times 10^{-9}$& $2\times 10^{-6}$\\
	$\Mw \ [\text{MeV}]$ & $782.59(4)$ & $782.59(4)$ & $782.60(4)$ & $782.63(2)$ & $782.63(2)$& $782.63(2)$\\
	$\Gw \ [\text{MeV}]$ & $8.70(6)$ & $8.70(6)$ & $8.68(6)$  & $8.67(3)$ & $8.67(3)$& $8.68(3)$\\
	$\Mphi \ [\text{MeV}]$& $1019.23(2)$ & $1019.23(2)$ & $1019.24(2)$ & $1019.21(1)$ & $1019.20(1)$& $1019.21(1)$\\
	$\Gphi \ [\text{MeV}]$ & $4.35(4)$ & $4.35(4)$ & $4.36(4)$  & $4.25(3)$ & $4.25(3)$& $4.25(3)$\\
	$c_\omega \ [\text{GeV}^{-1}]$ & $2.86(1)$ & $2.86(1)$ & $2.86(1)$ & $2.82(2)$ & $2.83(2)$& $2.82(2)$\\
	$c_\phi \ [\text{GeV}^{-1}]$ & $-0.395(3)$ & $-0.395(3)$ & $-0.395(3)$ & $-0.388(3)$ & $-0.388(3)$& $-0.389(3)$\\
	$c_{\omega'} \ [\text{GeV}^{-1}]$ & $-0.08(3)$ & $-0.09(5)$ & $0.10(5)$  & $-0.07(3)$ & $-0.07(4)$& $0.09(4)$\\
	$c_{\omega''} \ [\text{GeV}^{-1}]$ & $-0.86(6)$ & $-0.87(7)$ & $3.48(8)$ & $-0.85(6)$ & $-0.85(6)$& $3.42(8)$\\
	$c_1 \ [\text{GeV}^{-3}]$ & $-1.45(6)$ & $-1.45(7)$ & $-2.07(5)$ & $-1.42(6)$ & $-1.42(6)$& $-2.02(6)$\\
	$c_2 \ [\text{GeV}^{-3}]$ & $-0.60(9)$ & $-0.60(9)$ & $-1.83(5)$ & $-0.63(11)$ & $-0.62(11)$& $-1.80(5)$\\
	$c_3 \ [\text{GeV}^{-3}]$ & --- & $-0.08(6)$ & $-0.60(5)$ & --- & $-0.03(6)$& $-0.55(5)$\\
	$c_4 \ [\text{GeV}^{-3}]$ & --- & --- & $2.90(11)$ & --- & --- & $2.84(11)$ \\
	$10^4\times \xi$ & $1.9(7)$ & $1.9(7)$ & $1.9(7)$ & $1.3(5)$ & $1.3(5)$ & $1.4(5)$ \\
	$10^{10}\times a_\mu^{3\pi}|_{\leq 1.8\GeV}$ & $46.59(22)$ & $46.60(22)$ & $46.55(21)$ & $45.54(52)$ & $45.54(52)$ & $45.40(48)$\\ 
	\bottomrule
	\renewcommand{\arraystretch}{1.0}
	\end{tabular}
	\caption{Same as Table~\ref{tab:fits_combination}, but with $\Im C_p(q^2)\sim q^{-3}$ asymptotically.}
	\label{tab:fits_combination_asym}
\end{table}

\subsection[Extracting $\omega$ and $\phi$ masses]{Extracting $\boldsymbol{\omega}$ and $\boldsymbol{\phi}$ masses}
\label{sec:vector_meson_masses}

Our final result for the $\omega$ and $\phi$ parameters is
\begin{align}
\label{omega_phi_fit}
 \Mw&=782.63(3)(1)\MeV=782.63(3)\MeV,\notag\\
 \Gw&=8.71(4)(4)\MeV=8.71(6)\MeV,\notag\\
 \Mphi&=1019.20(2)(1)\MeV=1019.20(2)\MeV,\notag\\
 \Gphi&=4.23(4)(2)\MeV=4.23(4)\MeV,
\end{align}
with systematic errors derived as described in Sect.~\ref{sec:combined}. In the comparison to the PDG parameters~\cite{Tanabashi:2018oca}
\begin{align}
\label{omega_phi_PDG}
\Mw&=782.65(12)\MeV, & \Gw&=8.49(8)\MeV,\notag\\
\Mphi&=1019.461(16)\MeV, &\Gphi&=4.249(13)\MeV,
\end{align}
one needs to keep in mind that these parameters subsume radiative effects, where the expected corrections are worked out in App.~\ref{app:mass_shifts}.
For the $\phi$, the expectation is that the fit of the bare parameters should produce a mass lower by $0.26\MeV$ with only small corrections in 
the width, in perfect agreement with~\eqref{omega_phi_fit} and~\eqref{omega_phi_PDG}.   
In contrast, the situation for the $\omega$ is more ambiguous: the number in~\eqref{omega_phi_PDG} is dominated by the weighted average 
of extractions from $e^+e^-\to3\pi$ ($\Mw=782.68(9)(4)\MeV$~\cite{Akhmetshin:2003zn}, $\Mw=782.79(8)(9)\MeV$~\cite{Achasov:2003ir}),
$e^+e^-\to\pi^0\gamma$ ($\Mw=783.20(13)(16)\MeV$~\cite{Akhmetshin:2004gw}), and $\bar p p\to\omega\pi^0\pi^0$ ($\Mw=781.96(13)(17)\MeV$~\cite{Amsler:1993pr}). 
In view of the expected downward shift of $0.13\MeV$, our analysis thus supports the $3\pi$ number from~\cite{Achasov:2003ir}, while the agreement with the PDG average
is entirely coincidental.
As argued in~\cite{Colangelo:2018mtw}, the $\pi^0\gamma$ value is likely affected by an unphysical phase in the extraction, but our analysis shows that 
a similar effect does not occur in $3\pi$. Therefore, our analysis compounds the tension with the VP-subtracted $\omega$ mass as extracted from the $2\pi$ channel, 
$\Mw=781.68(9)(3)\MeV$~\cite{Colangelo:2018mtw}. 
Including the expected upward shift of $0.06\MeV$, the result for the width agrees with~\eqref{omega_phi_PDG} at the level of $1.6\sigma$, 
again consistent with earlier extractions from the $3\pi$ channel ($\Gw=8.68(23)(10)\MeV$~\cite{Akhmetshin:2003zn}, $\Gw=8.68(4)(15)\MeV$~\cite{Achasov:2003ir}).

\section{Consequences for the anomalous magnetic moment of the muon}
\label{sec:amu}

Our central result for the $3\pi$ contribution to HVP is
\beq
\label{final_result}
a_\mu^{3\pi}|_{\leq 1.8\GeV}=46.2(6)(6)\times 10^{-10}=46.2(8)\times 10^{-10},
\eeq
where the systematic errors are estimated as in Sect.~\ref{sec:combined}.
As a cross check we have also performed a fit to the data combination of~\cite{Keshavarzi:2018mgv}
instead of the data directly, leading to almost the same central value
\beq
\label{result_KNT}
a_\mu^{3\pi}|_{\leq 1.8\GeV}=46.1(6)(8)\times 10^{-10}=46.1(1.0)\times 10^{-10},
\eeq
with slightly larger uncertainties. The latter is likely related to the fact that 
although, as expected, we had to remove two bins (\# 49 and \# 52) corresponding to (nearly) vanishing cross sections to have the fit iteration converge,
no further changes were applied to the combination, so that some of the potentially problematic points we identified in~\cite{Akhmetshin:2006sc}
could still impact the fit.
The final result~\eqref{final_result} agrees well with $a_\mu^{3\pi}|_{\leq 1.8\GeV}=46.2(1.5)\times 10^{-10}$~\cite{Davier:2017zfy},
besides a corroboration of the central value the QCD constraints also allow for a reduction of the uncertainty. 
The difference to $a_\mu^{3\pi}|_{\leq 1.8\GeV}=47.7(9)\times 10^{-10}$~\cite{Keshavarzi:2018mgv} is mainly due to the interpolation applied to the data. We reproduce
the central value with a linear interpolation of the bins of~\cite{Keshavarzi:2018mgv}, while higher-order interpolations, as well as the dispersive fit~\eqref{result_KNT},
move the central value towards~\eqref{final_result}.\footnote{We thank B.~Malaescu and D.~Nomura for confirming that the choice of interpolation indeed 
explains the bulk of the difference between~\cite{Davier:2017zfy,Keshavarzi:2018mgv}.} 
Our analysis does not support values as low as
$a_\mu^{3\pi}|_{\leq 2.0\GeV}=44.3(1.5)\times 10^{-10}$~\cite{Jegerlehner:2017gek}, which is based on a Breit--Wigner description of $\omega$ and $\phi$. 
Finally, we remark that for the threshold region we find a value
$a_\mu^{3\pi}|_{\leq 0.66\GeV}=0.019\times 10^{-10}$ nearly twice as large as the estimate from~\cite{Hagiwara:2003da}
based upon a combination of the Wess--Zumino--Witten action and vector meson dominance~\cite{Kuraev:1995hc,Ahmedov:2002tg}.
Indeed, it was observed in~\cite{Hagiwara:2003da} that this model underestimates the lowest-energy data points.

In combination with the $2\pi$ channel from~\cite{Colangelo:2018mtw} we obtain for the HVP contribution that has been evaluated 
imposing analyticity and unitarity constraints
\beq
a_\mu^{2\pi}|_{\leq 1.0\GeV}+a_\mu^{3\pi}|_{\leq 1.8\GeV}=\big[495.0(2.6)+46.2(8)\big]\times 10^{-10}=541.2(2.7)\times 10^{-10},
\eeq
which covers nearly $80\%$ of the total HVP integral. In combination with the remaining contributions from~\cite{Davier:2017zfy,Keshavarzi:2018mgv}, we estimate
for the full leading-order HVP
\beq
\label{result_HVP}
a_\mu^\text{HVP}=692.3(3.3)\times 10^{-10}.
\eeq
Even assuming for the HLbL contribution a value as conservative as $a_\mu^\text{HLbL}=10(4)\times 10^{-10}$, this result thus reaffirms the $(g-2)_\mu$ anomaly at the 
level of $3.4\sigma$.\footnote{This estimate includes QED~\cite{Aoyama:2017uqe}, electroweak~\cite{Gnendiger:2013pva}, next-to-leading-order HVP~\cite{Keshavarzi:2018mgv},
next-to-next-to-leading-order HVP~\cite{Kurz:2014wya}, and next-to-leading-order HLbL~\cite{Colangelo:2014qya} contributions. Note that for $(g-2)_\mu$ it does
not matter if $\alpha$ is taken from $(g-2)_e$~\cite{Hanneke:2008tm} or Cs interferometry~\cite{Parker:2018vye}, 
but the tension between the two at the level of $2.5\sigma$ may by itself provide a first glimpse of physics beyond the SM~\cite{Davoudiasl:2018fbb,Crivellin:2018qmi}.}

\section{Summary}
\label{sec:summary}

We have presented a detailed analysis of the $3\pi$ contribution to HVP, including constraints from analyticity and unitarity as well as the
low-energy theorem for the $\gamma^*\to3\pi$ amplitude. Similarly to the $2\pi$ analysis of the pion vector form factor~\cite{Colangelo:2018mtw}, 
the main motivations are, first, to see if a global fit subject to these constraints reveals inconsistencies in the data, and, second,
derive the corresponding error estimate for the contribution to $(g-2)_\mu$. Given that this method is complementary to a direct
integration of the data, where potential inconsistencies are addressed by a local error inflation, such global fits 
that incorporate general QCD constraints should increase the robustness of the SM prediction.

We find that most data sets can be fit satisfactorily with our dispersive representation, the exception being several points 
above the $\phi$ resonance from~\cite{Akhmetshin:2006sc}. Fortunately, 
the impact on the final HVP integral is minimal,
due to the suppression of the cross section in this region, which could also enhance the relative importance of 
systematic effects in the data. Otherwise, in the $3\pi$ channel there is no tension between two high-statistics data sets, such as BaBar and KLOE in the $2\pi$ case,
but the scale factor of the global fit, indicating overall consistency of the data base, is actually larger than in $2\pi$.
In addition, the main contribution, from the $3\pi$ cross section in the vicinity of the $\omega$, is dominated by a single experiment~\cite{Achasov:2003ir}.
For these reasons, a new high-statistics low-energy measurement in the $3\pi$ channel would be a highly welcome addition to the data base.

The central outcome of our study is~\eqref{final_result}
\beq
a_\mu^{3\pi}|_{\leq 1.8\GeV}=46.2(6)(6)\times 10^{-10}=46.2(8)\times 10^{-10}.
\eeq
Together with the $2\pi$ channel from~\cite{Colangelo:2018mtw}, the two most important low-energy channels have now been scrutinized including
analyticity and unitarity constraints, covering nearly $80\%$ of the HVP integral. Depending on the assumptions for HLbL scattering,
the current tension is thus confirmed at the known level around $3.5\sigma$.
To make further progress, especially in view of the ultimate 
precision expected at the Fermilab $(g-2)_\mu$ experiment, new data input in particular in the $2\pi$ channel is critical.

Finally, our analysis exacerbates a tension emerging between the $2\pi$ and $3\pi$ channels, that is, the extraction of the $\omega$ mass.
In the $2\pi$ channel the $\omega$ only contributes via an isospin-violating effect, $\rho$--$\omega$ mixing, but due to the increased statistics the sensitivity
is not much below that of the $3\pi$ channel. Yet, the $\omega$ mass extracted from the $2\pi$ channel is substantially lower than the one extracted from $3\pi$. 
Currently, we are aware of neither a systematic effect in experiment nor an issue with the theoretical extraction that could resolve the tension. 
Besides improving the HVP contribution to $(g-2)_\mu$, new data could shed light on this puzzle as well.

\acknowledgments
We thank M.~Davier, A.~Keshavarzi, B.~Malaescu, D.~Nomura, T.~Teubner, and Z.~Zhang for numerous useful discussions. 
B.-L.~H. is grateful to S.~Holz and S.~Ropertz for discussions of avail. 
The dispersive representation of the $\gamma^*\to3\pi$ amplitude was
developed in collaboration with S.~P.~Schneider in the context of the pion transition form factor, 
generalizing earlier work on $\omega,\phi\to3\pi$
---we gratefully acknowledge his contribution.
We would further like to thank G.~Colangelo and P.~Stoffer, the present paper is very close in spirit 
to the $2\pi$ case~\cite{Colangelo:2018mtw} and relies strongly on the strategies developed therein.
In particular, we thank P.~Stoffer for valuable comments on the manuscript.
Finally, we thank C.~Hanhart for collaboration leading to the results given in App.~\ref{app:mass_shifts}.
Financial support was provided by
the DFG (CRC 110,
``Symmetries and the Emergence of Structure in QCD'')
and the DOE (Grant No.\ DE-FG02-00ER41132).

\appendix

\section{Electromagnetic mass shifts}
\label{app:mass_shifts}

The separation of VP from the full cross section affects the $\omega$ and $\phi$ pole parameters because the VP function itself
involves the corresponding poles, only suppressed by $e^2$. The size of the expected shifts can be analyzed analytically 
in a Bethe--Salpeter multi-channel approach~\cite{Hanhart:2012wi,inprep}. For instance, the $\omega$ contribution to the VP function $\Pi(s)$ becomes
\beq
\label{Pi_omega_poly}
\Pi_{\omega}(s)=\frac{e^2s}{\gwg^2}\frac{1}{s-\Mw^2+i\Mw\Gw},
\eeq
where the coupling is related to the two-electron width
$\Gamma_{\omega\to e^+e^-}=e^4\Mw/(12\pi \gwg^2)$,
i.e.\ $\gwg=16.7(2)$~\cite{Tanabashi:2018oca}. Expanding around the shifted pole parameters, one finds the relation
\beq
\label{omega_parameters_radiative_corrections}
\Mwb=\bigg(1+\frac{e^2}{2\gwg^2}\bigg)\Mw+\Order(e^4),\qquad \Gwb=\bigg(1+\frac{e^2}{2\gwg^2}\bigg)\Gw+\Order(e^4),
\eeq
where $\Mwb$ and $\Gwb$ include the effects of VP, while $\Mw$ and $\Gw$ should be identified with the fit parameters in~\eqref{omega_phi_fit}.
Numerically, \eqref{omega_parameters_radiative_corrections} implies 
\beq
\Delta\Mw=\Mwb-\Mw=0.13\MeV,\qquad \Delta\Gw=\Gwb-\Gw=1.4\keV.
\eeq
$\Mw$ is thus expected to be about $0.13\MeV$ lower than in PDG conventions, while the effect on the width due to $\Pi_{\omega}$ is negligible.
The same argument for the $\phi$ produces a mass shift 
\beq
\Delta \Mphi=\bar M_\phi-\Mphi=\frac{e^2}{2g_{\phi\gamma}^2}\Mphi=0.26\MeV.
\eeq

Finally, for the $\omega$ width there is an additional effect due to $\rho$--$\omega$ mixing, i.e., a higher-order effect enhanced by the small mass difference between $\omega$ and $\rho$.
In a vector-meson-dominance approximation for the $\rho$ we find the relation
\beq
\Delta\Gw=\frac{e^2}{2\gwg^2}\Gw+\frac{\Mw^2}{\Gr-\Gw}\frac{e^2}{\grg^2}\bigg(\frac{e^2}{\gwg^2}-2\eps_\omega\bigg)=-0.06\MeV,
\eeq
with mixing parameter $\eps_\omega\sim 2\times 10^{-3}$~\cite{Colangelo:2018mtw},
and by comparing fits with and without VP we verified that this indeed describes well the observed shift in the $\omega$ width.

\section{Estimate of the $\boldsymbol{F}$-wave contribution}
\label{app:F_waves}

For $q^2=0$~\cite{Hoferichter:2017ftn} and $q^2=\Mw^2$~\cite{Niecknig:2012sj} the impact of $F$-waves on the $\gamma^*\to3\pi$ amplitude was shown to be completely negligible below the $\rho_3(1690)$ resonance,
but since we consider virtualities up to $\sqrt{q^2}=1.8\GeV$ one may ask the question whether the impact of these resonant $F$-waves can still be ignored.  
There is little phenomenological information on the $\rho_3\pi\gamma^*$ coupling besides the $\rho_3\to\pi\omega$ branching ratio. However, the fact that the corresponding $\omega$-dominance estimate 
from~\cite{Hoferichter:2017ftn} is in line with preliminary results from COMPASS~\cite{Seyfried} suggests that at least within $[0,\Mw^2]$ the $q^2$-dependence should be approximately described by $a(q^2)$.
Here, we estimate a potential $F$-wave contribution by assuming that this approximation remains meaningful up to $\sqrt{q^2}=1.8\GeV$. 

The decomposition of the amplitude including $F$-waves becomes~\cite{Niecknig:2012sj}
\begin{align}
\F(s,t,u;q^2)&=\F(s;q^2)+\F(t;q^2)+\F(u;q^2)\notag\\
&+ P_3'(z_s)\G(s;q^2) + P_3'(z_t)\G(t;q^2) + P_3'(z_u)\G(u;q^2),
\end{align}
where the scattering angles follow by permuting the Mandelstam variables in~\eqref{kinematics} accordingly. To estimate the $\rho_3$ contribution, we first establish the connection 
to a narrow-resonance approximation of the $P$-wave
\beq
\F_\rho(s;q^2)=a(q^2)\frac{M_\rho^2}{M_\rho^2-s},
\eeq
with $M_\rho^2\to M_\rho^2-iM_\rho\Gamma_\rho$ in the decay region. We can then estimate the $\rho_3$ contribution as
\begin{align}
\G_{\rho_3}(s;q^2)&=a(q^2)\frac{M_{\rho_3}^2}{M_{\rho_3}^2-s}C_{\rho_3}\frac{\sigma^2_\pi(s)\lambda(q^2,\mpi^2,s)}{\Mw^4},\notag\\
C_{\rho_3}&=\frac{\pi^2g_{\rho_3\pi\pi}g_{\rho_3\pi\omega}\Mw^4}{5g_{\omega\gamma}M_{\rho_3}^2},\qquad |C_{\rho_3}|\sim 1\times 10^{-3},
\end{align}
where the coupling constants are set to the values from~\cite{Hoferichter:2017ftn}. 
Numerically, we find that the interference between $P$- and $F$-waves gives a correction around $1\%$ at $\sqrt{q^2}=1.8\GeV$, while the pure $F$-wave contribution
is suppressed by another two orders of magnitude. 
These results confirm the expectation that the $\rho_3(1690)$ should not become relevant until well above the threshold $M_{\rho_3}+\mpi\sim 1.83\GeV$ where the decay becomes possible.

\end{document}